\documentclass{elsart}
\usepackage{amsfonts}
\usepackage{amssymb}
\usepackage{graphicx}
\usepackage{algorithmic}
\newcommand{\ANDOP}{\textbf{and}}
\newcommand{\eff}{\ensuremath{E_p}}

\newcommand{\Vcubic}{\ensuremath{V_{\mathit{cubic}}}}
\newcommand{\Scubic}{\ensuremath{S_{\mathit{cubic}}}}
\newcommand{\SBCC}{\ensuremath{S_{\mathit{BCC}}}}
\newcommand{\SFCC}{\ensuremath{S_{\mathit{FCC}}}}
\newcommand{\Sslices}{\ensuremath{S_{\mathit{slices}}}}
\newcommand{\Scolumns}{\ensuremath{S_{\mathit{columns}}}}

\begin{document}

%-----------------------------------------------------------------
% Title and author definitions
%-----------------------------------------------------------------

% Uncomment for standard LaTex:
%\maketitle

% Uncomment for 'C' style journals:
%\authorrunninghead{M.A. Stijnman, R.H. Bisseling, G.T. Barkema}
%\titlerunninghead{Partitioning 3D space for parallel particle
%simulations}
%\title{ Partitioning 3D space for parallel many-particle
%simulations}
%\author{M.A. Stijnman, R. H. Bisseling}
% \affil{Mathematical Institute, Utrecht University,\\
% PO Box 80010, 3508 TA Utrecht, The Netherlands}
%\and
%\author{G. T. Barkema}
% \affil{Institute for Theoretical Physics, Utrecht University,\\
% Leuvenlaan 4, 3584 CE Utrecht, The Netherlands}

% Uncomment for Elsevier jourals:
\begin{frontmatter}
\title{Partitioning 3D space for parallel many-particle simulations}

\author{M.A. Stijnman \and R. H. Bisseling}
\address{Mathematical Institute, Utrecht University,\\
 PO Box 80010, 3508 TA Utrecht, The Netherlands}
\author{G. T. Barkema}
\address{Institute for Theoretical Physics, Utrecht University,\\
 Leuvenlaan 4, 3584 CE Utrecht, The Netherlands}
 
% Uncomment for 'C' style jornals:
% \keywords{
% Uncomment for Elsevier journals:

% Uncomment for 'C' style jornals:
%\abstract{
% Uncomment for standard laTeX and Elsevier journals:
\begin{abstract}
In a common approach for parallel processing applied to
simulations of many-particle systems with short-ranged interactions and uniform
density, the simulation cell is partitioned into domains of equal shape and
size, each of which is assigned to one processor. We compare the commonly used simple-cubic (SC) domain shape to domain shapes chosen as the Voronoi cells of BCC and FCC lattices. The latter two are found to result in superior partitionings with respect to communication overhead. Other domain shapes, relevant for a small number of processors, are also discussed. The higher efficiency with BCC and FCC partitionings is demonstrated in simulations of the sillium model for amorphous silicon.
% Uncomment for 'C' style jornals:
%}
% Uncomment for standard laTeX and Elsevier journals:
\end{abstract}

% Uncomment for 'C' style journals:
%\begin{article}
%\draft

% Uncomment for Elsevier journals:
\end{frontmatter}

%-----------------------------------------------------------------
% Start of article text
%-----------------------------------------------------------------

\section{Introduction}\label{sec:intro}
Realistic simulations of molecular dynamics and other dynamic
many-particle systems demand increasingly larger models. Calculations
on these large models can be distributed over several processors of a
parallel computer to improve performance. An excellent review of the
state-of-the-art of parallel atomistic simulations has recently been
published by Heffelfinger~\cite{heffelfinger00}. According to this
work, and to the best of our knowledge, spatial decomposition of the
simulation cell is done almost exclusively by partitioning into cubic
domains of equal size, each of which is assigned to a
processor. Exceptions to this rule are earlier work by Esselink and
Hilbers~\cite{esselink93} and Chynoweth et al.~\cite{chynoweth91}, who
use a 2D decomposition motivated by the square mesh topology of their
parallel machine. In case of density fluctuations, load imbalance
between the processors might occur; here, we limit ourselves to
homogeneous systems with a uniform density, such as bulk materials or
liquids. Other methods are necessary for heterogeneous systems such as
proteins in vacuum or stellar systems. In homogeneous many-particle
systems, the major source of inefficiency inherent to the domain
decomposition approach lies in the fact that particles interact over
some distance, so that in particular particles near the surface of
these domains interact with particles in neighbouring domains. These
particles near the surface thus cause communication with neighbouring
processors, redundant calculations, or both. For brevity, we call this
entire extra work the communication overhead. In the case that the
interaction range is much smaller than the lateral size of the
domains, the communication overhead will roughly scale with the
surface area of the domain. Hence, the optimal domain shape for
many-particle systems with a uniform density and a short-range
potential is a space-filling shape with minimal surface-to-volume
ratio. 

This paper is organized as follows. First, we explore domain shapes
that are derived from simple-cubic (SC), body-centered-cubic (BCC) and
face-centered-cubic (FCC) lattices. We determine their properties with respect
to their use in parallel processing and discuss several implementation
details. We then apply these domain shapes in a representative
many-particle simulation: the {\em sillium\/} model of amorphous
silicon, as proposed by Wooten, Winer and Weaire. Finally, we present
our conclusions. 

\section{Domain shapes}\label{sec:dist}
In this section, several domain shapes are discussed, regarding their
properties relevant for parallel processing. All lengths and distances
are measured in fractions of the system to be simulated, which thus by
definition has unit length edges. The domains assigned to each
processor are equal in shape and size, and consequently have a volume
of $1/p$ where $p$ is the number of processors. The following
discussion assumes a cubic simulation cell, but extension to other
regular-shaped simulation cells is straightforward. The interaction
range (distance over which particles exert forces) equals $r_c$, where
$r_c \ll 1$. Relevant for our purpose is the volume of the {\em
halo\/}: the region outside the domain but within a distance
$r_c$. Particles in this halo interact with those inside the domain,
causing communication overhead. 

\subsection{SC partitioning}
The most straightforward three-dimensional division of a cube into
identical domains is a division into $p=k^3$ smaller cubes, with $k$ a
positive integer. The resulting cubic domains have an edge length of
$1/k$. The volume \Vcubic\ of the halo with radius $r_c$ around each
domain equals 
\begin{equation}
\Vcubic=
%\frac{1}{k^3}+
\frac{6r_c}{k^2}+
\frac{3\pi r_c^2}{k}+
\frac{4}{3}\pi r_c^3.
\label{vcubiceq1}
\end{equation}

The first term is dominant and equal to $r_c$ times the surface area
of the domain. The second and third terms correspond to the extra
volume of the halo located near the edges and corners of the domain,
respectively. In simulations with short-range interactions as
discussed here, usually $r_c k < 1$, so that these terms are small
compared to the first.

In the limit of very short-range interaction our problem reduces to
the well known Kelvin problem, which is to find a partitioning of
space with minimal surface area. Kelvin~\cite{kelvin1887} conjectured
that the optimal solution is the Voronoi cell of the BCC lattice,
slightly curved to satisfy Plateau's rules~\cite{plateau1873}, but
Weaire and Phelan~\cite{weaire94} produced an even better
partitioning, based on two different cells, and related to the
$\beta$-tungsten structure. We limit ourselves to proposing
partitionings that can be shown to be better than SC and that can be
implemented in a relatively simple way. 

In the case of SC, the surface area is equal to
\begin{equation}
\Scubic=\frac{6}{k^{2}}=\frac{6}{p^{\frac{2}{3}}}.
\label{squbiceq}
\end{equation}

\subsection{BCC partitioning}
\begin{figure}
\begin{center}
\begin{tabular}{cc}
\includegraphics[scale=0.7]{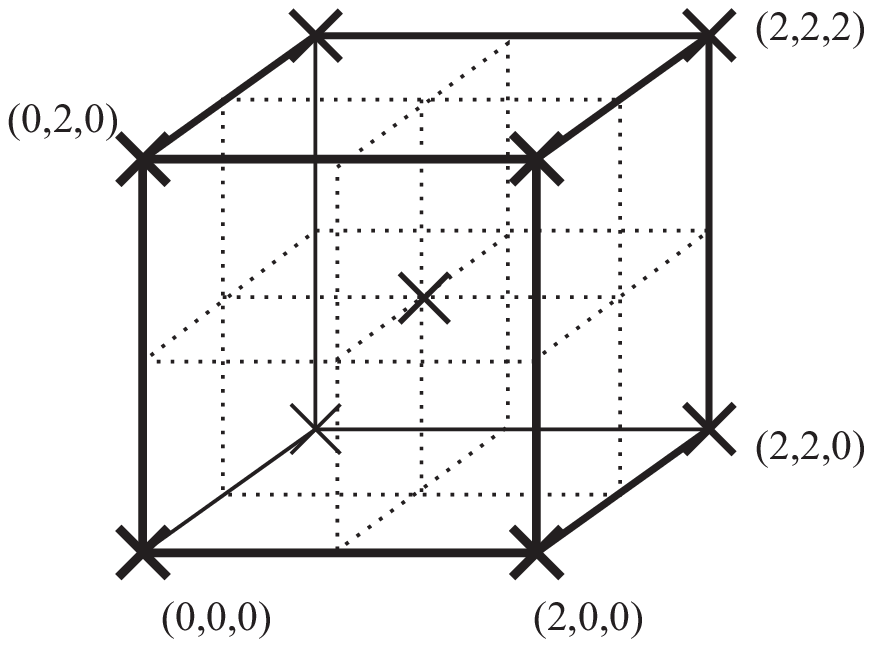} &
\includegraphics[scale=0.7]{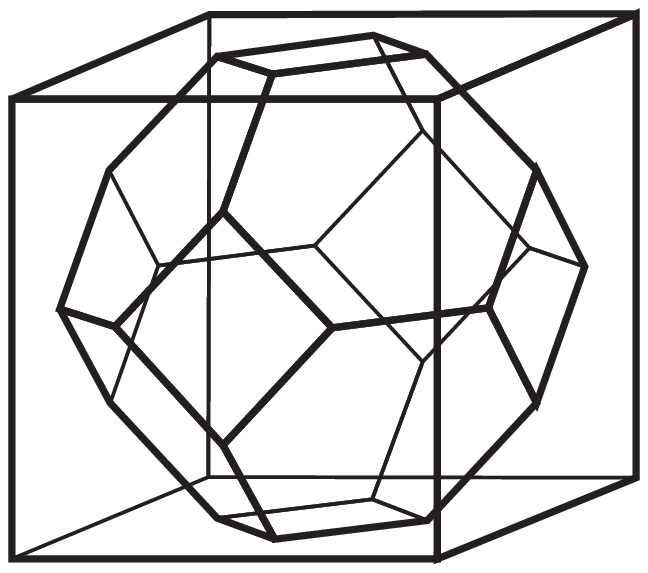}\\
%\textbf{(a)} & \textbf{(b)}
(a) & (b)
\end{tabular}
\end{center}
\caption{(a) The basic BCC lattice cell. Sphere centres are marked by
`$\times$'. (b) The BCC Voronoi cell, a truncated octahedron.}
\label{fig:bcc}
\end{figure}

Given that we strive for a small surface-to-volume ratio, it is
natural to investigate sphere packings.  In one of the better sphere
packings, the spheres are located on the sites of a body-centred-cubic
(BCC) lattice, with spheres at the corners and the centres of cubic
cells. The BCC unit cell is displayed in Figure~\ref{fig:bcc}(a). Each
unit cell adds two sphere centres to the lattice, as only one corner
point is part of the unit cell; the other seven corner points are
considered part of neighbouring cells. By repeating this unit cell the
BCC lattice is generated. The lattice is then rescaled, such that the
length of the edges of a unit cell becomes $\frac{1}{k}$.

The domain of a processor is formed by the Voronoi cell of a lattice
point, i.e., the space closest to that point. The model cube can now
be divided into $p=2k^3$ Voronoi cells, as generated by the BCC
lattice. It turns out that each Voronoi cell is a truncated
octahedron, as shown in Figure~\ref{fig:bcc}(b). This is also the shape
that Kelvin proposed as a solution to the Kelvin problem. 

Each truncated octahedron generated by the BCC lattice fits into a
cube with edge length $\frac{1}{k}$. Each of the six square faces has
a surface area of $\frac{1}{8k^2}$ and each of the
eight hexagonal faces has
a surface area of $\frac{3\sqrt{3}}{16k^2}$. This results in a
total surface area of
\begin{equation}
\SBCC=\frac{6\sqrt{3}+3}{4k^2},
\end{equation}
which, after substitution of $k=(p/2)^{\frac{1}{3}}$, gives
\begin{equation}
\SBCC=\frac{6\sqrt{3}+3}{2^{\frac{4}{3}}p^\frac{2}{3}}
\approx\frac{5.3147}{p^\frac{2}{3}}.
\end{equation}
This is over eleven percent better than for SC.

\subsection{FCC partitioning}
\begin{figure}
\begin{center}
\begin{tabular}{cc}
\includegraphics[scale=0.7]{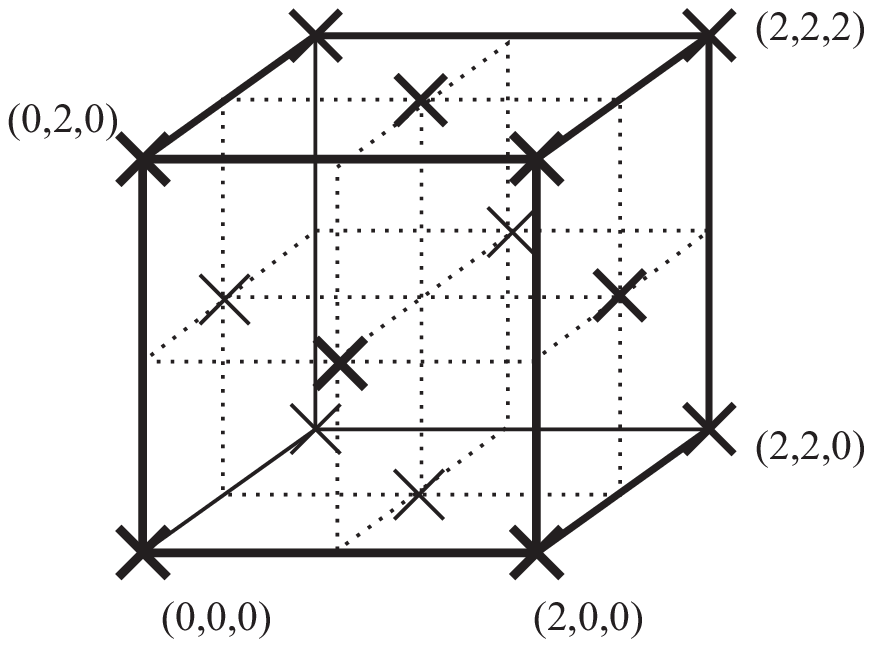} &
\includegraphics[scale=0.7]{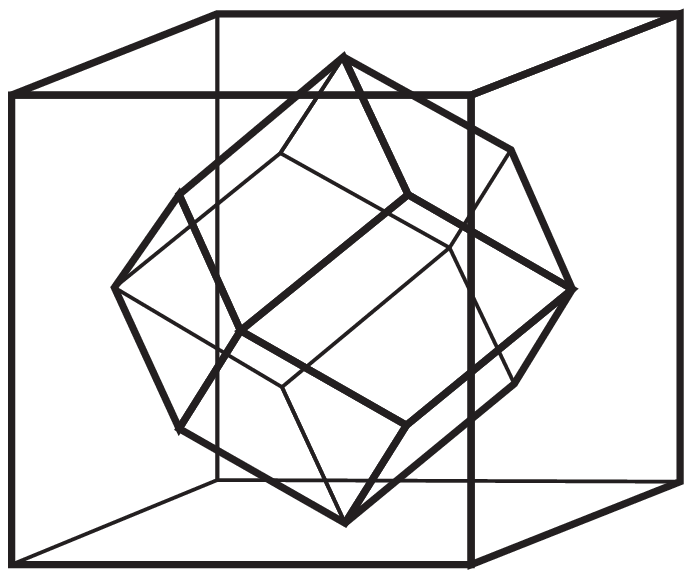}\\
%\textbf{(a)} & \textbf{(b)}
(a) & (b)
\end{tabular}
\end{center}
\caption{(a) The FCC basic lattice cell. Sphere centres are marked
by
`$\times$'. (b) The FCC Voronoi cell, a rhombic dodecahedron, 
translated to the centre of the cube for reference.}
\label{fig:fcc}
\end{figure}

In one of the optimal dense sphere packings, the spheres are placed at
sites of a face-centered-cubic (FCC) lattice, with spheres at the
corners of cubic cells and at the centres of the faces. The FCC unit
cell is shown in Figure~\ref{fig:fcc}(a). The corresponding Voronoi cell
is a rhombic dodecahedron, as shown in Figure~\ref{fig:fcc}(b). Each
cubic unit cell adds four points to the lattice, so with this
partitioning $p=4k^3$ processors can be used. After the lattice is
rescaled such that each unit cell has an edge of length $\frac{1}{k}$,
the Voronoi cell can be considered as being made up of a cube with
edge length $\frac{1}{2k}$ and six pyramids with height
$\frac{1}{4k}$, each covering one face of the small cube. The surface
area of the rhombic dodecahedron equals 24 times the surface area of
one of the triangular faces of the pyramid, which is
$\frac{1}{8\sqrt{2}k^2}$. The surface area of the domain in the FCC
partitioning therefore equals
\begin{equation}
\SFCC=\frac{3}{\sqrt{2}k^2},
\end{equation}
which, after substitution of $k=\sqrt[3]{p/4}$, yields
\begin{equation}
\SFCC=\frac{3\cdot 2^\frac{5}{6}}{p^\frac{2}{3}}
\approx\frac{5.3454}{p^\frac{2}{3}},
\end{equation}
which is slightly more than for BCC, but still almost eleven
percent better than for SC.

Most parallel computers are equipped with $p=2^q$ processors. With the
three partitionings presented above, we can now use $p=k^3$, $p=2k^3$
and $p=4k^2$ processors, which includes all powers of two.  This means
that most parallel computers can be used to their full potential.

\subsection{Domain shapes with few processors}
Two other simple partitionings exist that
have not been mentioned
yet. The first is the partitioning into slices with dimensions $1\times
1\times \frac{1}{k}$.  Each slice has two sides with unit surface area
where communication overhead is generated,

\begin{equation}
\Sslices=2.
\end{equation}
The domain surface area does not decrease with an increasing number of
processors, in contrast to the three-dimensional partitionings
discussed above. For this reason, this domain shape is only useful
when $p$ is small.  For $p=2$, partitioning into slices is better than
BCC partitioning (with area $2$ vs. $3.35$) and for $p=4$ it is better
than FCC partitioning (with area $2$ vs. $2.12$), but it is worse in
all other cases where one of the SC, BCC, or FCC partitionings is
applicable.

Another partitioning is that into columns with dimensions
$1\times\frac{1}{k} \times \frac{1}{k}$. Each column has a surface
area of
\begin{equation}
\Scolumns=\frac{4}{p^{\frac{1}{2}}}. 
\end{equation}
For $p=4$ this is as efficient as using slices. Three-dimensional
partitioning is better at higher $p$.

\section{Practical implementation issues}\label{sec:impl}
In implementations, one needs an efficient procedure to determine the
processor to which a particle with coordinates $(x,y,z)$ is
assigned. Note that we can break ties arbitrarily in case particles
are located exactly 
on the boundary of two domains
(or within a distance corresponding to machine precision),
since this event is very unlikely to occur. With SC
partitioning for $p=k^3$ processors, the processor number $s$ can then
be found by: 
\begin{equation}
s=\lfloor kx\rfloor + k\lfloor ky\rfloor + k^2\lfloor kz\rfloor,
\end{equation}
where $\lfloor x\rfloor$ is the largest integer smaller than or equal to $x$. 

To assign processor numbers to domains with the BCC shape, it is
helpful to note that a BCC lattice consists of two simple-cubic
lattices, shifted with respect to each other over a vector
$(\frac{1}{2k},\frac{1}{2k},\frac{1}{2k})$. One simply determines the
nearest site in each of the two sublattices, compares the two, and
takes the nearest. Here, one can take Manhattan distances, defined by
$\|(x,y,z)\|_1= |x|+|y|+|z|$, because these are cheaper to compute than
Euclidean distances. Since the sum of the Manhattan distances
of an arbitrary point to the two nearest sites equals $\frac{3}{2k}$,
the nearest one is located at a Manhattan distance of less than
$\frac{3}{4k}$. The procedure to calculate the processor number for
the BCC partitioning is outlined in the pseudo-code below, where
`$[x]$' denotes the integer nearest to $x$ and `$\bmod$' denotes the
modulo operator (needed to wrap around the periodic boundaries):

\vspace{5mm}
\begin{algorithmic}
\STATE $D=|kx-[kx]|+ |ky-[ky]|+ |kz-[kz]|$
\IF{$D<\frac{3}{4}$}
\STATE $s=([kx]\bmod k) + k([ky]\bmod k) + k^2([kz]\bmod k)$
\ELSE
\STATE $s=k^3 + \lfloor kx\rfloor + k\lfloor ky\rfloor + k^2\lfloor
kz\rfloor$
\ENDIF
\end{algorithmic}

For the FCC lattice, it is helpful to note that it can be obtained
from a cubic superlattice by removing all grid points for which the
total of the coordinates is odd. First, we determine the nearest grid
point of this cubic superlattice. If the sum of the coordinates of
that grid point is even, that point was the nearest FCC lattice
site. If the sum is odd, the closest lattice point can be found by
rounding one of the coordinates in the `wrong' direction, i.e., in the
opposite direction of the nearest integer coordinate; the coordinate
that should be rounded `wrongly' is the one with the largest rounding
error (and hence the smallest error in the opposite direction). Using
the notation $]x[$ for rounding `wrongly' (in contrast to $[x]$ for
usual rounding), defined by the relation $[x]+]x[=\lfloor x \rfloor +
\lceil x\rceil$, this procedures thus becomes:

\vspace{5mm}
\begin{algorithmic}
\STATE{$(P_x,P_y,P_z)=([2kx],[2ky],[2kz])$}
\IF{$P_x+P_y+P_z$ is odd}
\IF{$|2kx-P_x|>|2ky-P_y|$ \ANDOP\ $|2kx-P_x|>|2kz-P_z|$} 
\STATE $P_x=]2kx[$
\ELSIF{$|2ky-P_y|>|2kz-P_z|$}
\STATE $P_y=]2ky[$
\ELSE
\STATE $P_z=]2kz[$
\ENDIF
\ENDIF
\STATE $(P_x,P_y,P_z)= (P_x\bmod 2k,P_y\bmod 2k,P_z\bmod 2k)$
\STATE $s=P_x + 2k P_y + 4k^2 \lfloor P_z/2 \rfloor$
\end{algorithmic}

\section{Related partitionings}\label{sec:reldist}

FCC is only one of the optimal sphere packings; another one is
hexagonal-close-packed (HCP). The Voronoi cells of HCP have the same
volume and surface area as those of FCC. The HCP packing, however, is
not derived from a cubic grid, so in an implementation it is more
difficult to assign processor numbers to particles.

The FCC partitioning can be used as the basis for yet another
partitioning, which we call 6FCC. The FCC cell can be further
subdivided into six (nonregular) octahedra, so $p=6\cdot 4k^3=24k^3$
processors can be used. This can be rewritten as $p=3k'^3$, with
$k'=2k$. Each octahedron consists of two pyramids with height
$\frac{1}{2k'}$ and base edges of length $\frac{1}{k'}$. The surface
area of this shape equals $\frac{4}{\sqrt{2}k'^2}$, which, after the
substitution $k'=(p/3)^{\frac{1}{3}}$, yields
\begin{equation}
S_{6\mathit{FCC}}=\frac{3^{\frac{2}{3}}4}{2^{\frac{1}{2}}p^{\frac{2}{3}}}
\approx \frac{5.8833}{p^{\frac{2}{3}}} .
\end{equation}
This is still slightly better than SC, but only really useful when a
parallel computer with $p=3k^3$ processors is used. The case with
$p=3$ is an exception, since using three slices is more
efficient. Note that this partitioning is equivalent to the FCC
partitioning with only the centres of the faces retained in the
lattice. 

A summary of recommended partitionings for typical numbers of
processors is given in Table \ref{tbl:summary}. This table illustrates
the added flexibility that is the result of having several different
partitioning methods in our toolbox. 

\begin{table}
\begin{center}
\begin{tabular}{rl@{\hspace{20mm}}rl}
\hline
$p$ & partitioning& $p$ & partitioning\\
\hline
1 & ---		& 24 & 6FCC		\\
2 & slices	& 27 & SC		\\
3 & slices	& 32 & FCC		\\
4 & slices	& 64 & SC		\\
8 & SC 		& 128 & BCC		\\
16 & BCC	& 256 & FCC		\\
\hline
\end{tabular}
\caption{Best partitioning for typical numbers of processors.} 
\label{tbl:summary}
\end{center}
\end{table}

\section{Application: amorphous silicon}\label{sec:appl}

We have applied SC, BCC, and FCC partitioning to the construction of
models of amorphous silicon, following the {\em sillium\/} model
proposed by Wooten, Winer and Weaire~\cite{www1,www2}, with recent algorithmic
improvements~\cite{highq}. This has produced the best models of
amorphous silicon that are available to date.

Within the sillium approach, an atomic configuration consists of the
coordinates of all $N$ atoms, together with a list of the $2N$ bonds
between them. The energy of a configuration is obtained from the
Keating potential~\cite{keating66}:
\begin{equation}
  E = \frac{3}{16} \frac{\alpha}{d^2} \sum_{\langle ij \rangle}
    \left( {\bf r}_{ij} \cdot {\bf r}_{ij}-d^2 \right)^2
    + \frac{3}{8} \frac{\beta}{d^2} \sum_{\langle jik \rangle}
    \left( {\bf r}_{ij} \cdot {\bf r}_{ik}+\frac{1}{3}d^2 \right)^2.
\end{equation}
Here, $\alpha$ and $\beta$ are the bond-stretching and bond-bending
force constants, respectively; $d=2.35$ \AA\ is the equilibrium Si-Si
bond length in the diamond structure. Usual values are $\alpha=2.965$
eV/\AA$^2$ and $\beta=0.285\; \alpha$.

The construction of a well-relaxed configuration starts from a
configuration in which atoms with random coordinates are four-fold
connected. This network is then relaxed through a sequence of many
proposed bond transpositions, accepted with the Metropolis acceptance
probability~\cite{Metrop} given by
\begin{equation}\label{Metrop}
  P = \min \left[ 1, \exp \left( \frac{E_b-E_f}{k_bT} \right)
\right],
\end{equation}
where $k_b$ is the Boltzmann constant, $T$ the temperature, and $E_b$ and
$E_f$ are the total {\em quenched\/} energies of the system before and after
the proposed bond transposition.

With the approach given above, and described in more detail in
Refs.~\cite{www1,www2}, Wooten, Winer and Weaire obtained 216-atom
structures of {\it a}-Si with a bond angle deviation as low as 10.9
degrees. A decade later, using the same approach but more computing
power, Djordjevi\'c, Thorpe and Wooten~\cite{djordjevic95} produced
larger (4096-atom) networks of even better quality, with a bond angle
deviation of 11.02 degrees for configurations without four-membered
rings and 10.51 degrees when these rings were
allowed~\cite{djordjevic95}. With some additional algorithmic
improvements, Barkema and Mousseau generated 1000-atom configurations
with a bond angle deviation of 9.2 degrees~\cite{highq}, and one
4096-atom configuration with a bond angle deviation of 9.89
degrees. Exploiting parallel processing, we have generated a
10,000-atom sample with a bond-angle deviation as low as 9.88
degrees and a 20,000-atom sample, used primarily for our benchmarking.
For a discussion of all structural and electronic properties
of the 10,000-atom sample, we refer to a forthcoming publication~\cite{stijnman};
here we focus on computational aspects.

In our parallel program, the model box containing 10,000 atoms was
divided using SC, BCC, and FCC partitionings, depending on the number
of processors used. Due to the three-body term in the Keating
potential, two extra layers of atoms are needed near the surface of
the domains. Communication of these extra atoms is determined by the
connectivity information in the model. It turned out that the amount
of communication needed can be well approximated by using a cutoff
distance of $1.3$ times the average bond length of 2.35\AA.
For the 10,000-atom and 20,000-atom sample, with box sizes 
57.5\AA\, and 72.6\AA, we find $r_c\approx 0.053$ 
and 0.042, respectively.

The program was tested for different values of $p$ on a Cray T3E
parallel computer, using the BSPlib communications
library~\cite{hill98},
and applied to the 20,000-atom sample.
As a simple performance metric, we take the efficiency \eff,
defined as 
\begin{equation}
\eff=\frac{T_1}{pT_p},
\end{equation}
where $T_p$ is the execution time of one iteration of the global
relaxation procedure on $p$ processors and $T_1$ is the time for one
processor without communication overhead.

\begin{figure}
\includegraphics{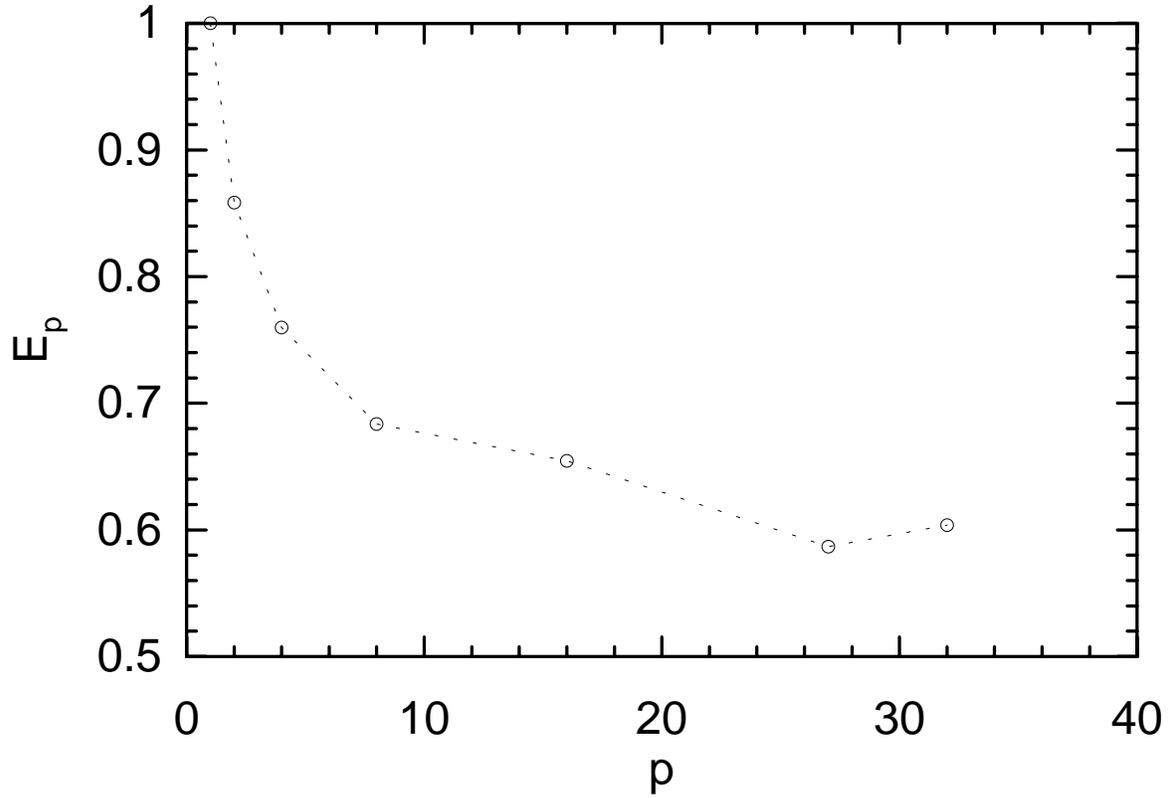}
\caption{Plot of the measured and predicted efficiency as a function
of the number of processors $p$. }
\label{fig:graph}
\end{figure}

The results of the efficiency measurements are shown in
Figure~\ref{fig:graph}. It is clear that in general the efficiency deceases
as $p$ increases. However, the sudden increase in efficiency when going
from $p=27$ to $p=32$ shows most clearly that FCC partitioning is
better than SC.
A similar effect can
also be observed, though less pronounced, at $p=16$ for BCC
partitioning.

As an illustration, the 20,000-atom sample was partitioned by the
different methods and the atoms in the inner region and the halos were
counted. Both the maximum and the average over the $p$ processors were
determined. 
(The maximum number of interior atoms determines
the computation time, whereas the maximum
number of halo atoms determines the communication time.) 
The results are listed in Table \ref{tbl:count}. Also
displayed is the ratio of the average number of halo atoms to the
average number of interior atoms, a metric that corresponds to the
surface-to-volume ratio. (The ratio of the averages is the
best measure of the effects studied, since it is less noisy
than the ratio of the maxima.) 
The ratios found explain the jumps in efficiency shown
in Figure~\ref{fig:graph}.

\begin{table}
\begin{center}
\begin{tabular}{rlrrrrr}
\hline
\multicolumn{1}{c}{$p$}& partitioning &
\multicolumn{2}{c}{interior} &
\multicolumn{2}{c}{halo} & ratio \\
%\hline
&&
max & avg & 
max & avg &  halo/interior \\
\hline
1 & ---	& 20000	& 20000	& 0	& 0	& 0.000 \\
2 & BCC	& 10009	& 10000	& 2382	& 2377	& 0.238 \\
4 & FCC	& 5019	& 5000	& 2156	& 2139	& 0.428 \\
8 & SC 	& 2551	& 2500	& 1546	& 1517	& 0.607 \\
16& BCC	& 1277	& 1250	& 904	& 874	& 0.699 \\
27& SC	& 765	& 741	& 723	& 706	& 0.953 \\
32& FCC	& 653	& 625	& 614	& 581	& 0.930 \\
\hline
\end{tabular}
\caption{The maximum and average number of atoms in the interior of
the processor domains and in the halos. Also listed is the ratio
between the average numbers of halo atoms and
interior atoms.}
\label{tbl:count}
\end{center}
\end{table}

\section{Conclusion}
We have proposed two space partitionings, based on Voronoi cells of
the BCC and FCC lattices, that can be used in parallel particle
simulations with uniform density and short-ranged interactions. The
advantage of these new partitionings is two-fold: (i) they reduce the
communication volume by about eleven percent compared to the commonly
used SC partitioning; (ii) they extend the range of possible processor
numbers, so that now we can use, among others, all powers of two as a
number of processors. These two partitionings are of
practical use, because they are almost as easy to implement as SC.

% Uncomment for 'C' style journals or standard LaTeX:
%\section*{Acknowledgment}
% Uncomment Elsevier journals:
\begin{ack}
We thank the Netherlands Computer Facilities foundation
and the High Performance Applied Computing center at Delft University
of Technology for providing access to a Cray T3E. 
% Uncomment for Elsevier journals:
\end{ack}

% Uncomment for 'C' style journals:
%\begin{references}
% Uncomment for standard LaTeX article or Elsevier journals:

% Uncomment for 'C' style journals:
%\end{references}
%\end{article}

\end{document}